\documentclass[aps,prd,twocolumn,superscriptaddress,floatfix]{revtex4-1}

\usepackage{graphicx}
\usepackage{amsmath}
\usepackage{amssymb}
\usepackage[colorlinks,citecolor=blue,linkcolor=blue,urlcolor=blue,anchorcolor=blue]{hyperref}
\allowdisplaybreaks
\makeatletter

\@addtoreset{equation}{section}
\makeatother
\newcommand\dx{{\rm d}}
\newcommand\p{\partial}
\newcommand\etal{{\it et~al.}}

\newcommand\ijmpa{Int. J. Mod. Phys. A}

\newcommand\jcap{J. Cosmol. Astropart. Phys.}
\newcommand\jhep{J. High Energy Phys.}

\newcommand\pdu{Phys. Dark Universe}
\newcommand\plb{Phys. Lett. B}

\begin{document}

\title{Revisiting scalar and tensor perturbations in a nonlocal gravity}
\author{S. X. Tian}
\email[]{tshuxun@whu.edu.cn}
\affiliation{School of Physics and Technology, Wuhan University, 430072 Wuhan, China}
\author{Zong-Hong Zhu}
\email[]{zhuzh@whu.edu.cn}
\affiliation{School of Physics and Technology, Wuhan University, 430072 Wuhan, China}
\affiliation{Department of Astronomy, Beijing Normal University, 100875 Beijing, China}
\date{\today}
\begin{abstract}
  Nonlocal RT gravity is a successful modified gravity theory, which not only explains the late-time cosmic acceleration but also behaves well in the solar system. Previous analysis generally assumes the auxiliary field $S_i$ vanishes at the cosmic background. However, we find the background $S_i$ is proportional to $a^2$ with the expansion of the universe. Then we discuss the influence of the nonzero background $S_i$ on the cosmic background evolution, the scalar and tensor perturbations. We find the cosmic background evolution is independent of $S_i$, and the influence of the nonzero background $S_i$ on the weak field limit at solar system scales is negligible. For the tensor perturbation, we find the only possible observable effect is the influence of nonzero background $S_i$ on the LIGO gravitational wave amplitude and also luminosity distance. Future high redshift gravitational wave observations could be used to constrain the background value of $S_i$.
\end{abstract}
\maketitle

\section{Introduction}\label{sec:01}
Nonlocal RT gravity is proposed by \cite{Maggiore2014a} to explain the late-time cosmic acceleration. Unlike the classical way of modifying gravity, \cite{Maggiore2014a} did not write down the Lagrangian but directly proposed the modified field equation
\begin{equation}\label{eq:1.1}
  G_{\mu\nu}-\frac{m^2}{3}(g_{\mu\nu}\Box^{-1}R)^\textrm{T}=\kappa T_{\mu\nu}.
\end{equation}
where the constant $m=\mathcal{O}(H_0/c)$. The dimension of $m$ is ${\rm length}^{-1}$. Note that until now we still do not know what kind of the Lagrangian corresponds to the above field equation. For the cosmological applications of the nonlocal RT gravity, \cite{Maggiore2014a,Foffa2014} analyzed the cosmological background evolution and \cite{Dirian2014,Dirian2015,Dirian2016} gave detailed cosmological perturbation analyses with a modified CLASS Boltzmann code. The observational constraints obtained in \cite{Dirian2014,Dirian2015,Dirian2016} show that the nonlocal RT gravity and  $\Lambda$CDM model perform equally well in cosmology. For the performance of the nonlocal RT gravity at solar system scales, \cite{Dirian2014,Nesseris2014} shows the weak field limit of the nonlocal RT gravity gives the Poisson equation, $\Psi=\Phi$ and $G={\rm const.}$, which means this theory can explain the dynamics of the solar system as general relativity does. This is an very important property as \cite{Belgacem2019a,Tian2019a} pointed out most of the nonlocal gravity theories cannot explain the dynamics of the solar system, e.g., the original Deser-Woodard theory \cite{Deser2007}, nonlocal RR gravity \cite{Maggiore2014b}, nonlocal Gauss-Bonnet gravity \cite{Capozziello2009} and the scalar-tensor nonlocal gravity \cite{Tian2018}. But note that the nonlocal RT gravity is not the only nonlocal theory that can explain the solar system dynamics as discussed in \cite{Giani2019}.

In order to solve Eq. (\ref{eq:1.1}), one need to introduce the auxiliary fields $U$ and $S_\mu$ as we presented in Sec. \ref{sec:02}. Previous works about the nonlocal RT gravity assumed $S_i=0$ in the cosmic background (see \cite{Maggiore2014a,Foffa2014,Dirian2014,Dirian2015,Dirian2016,Nesseris2014} for examples). However, as shown in Sec. \ref{sec:02}, the value of the background $S_i$ is proportional to $a^2$ with the expansion of the universe. Thus it is unreasonable to assume the background $S_i$ equals to 0 in the relevant analyses for the nonlocal RT gravity. This is our motivation to reanalyze the scalar and tensor perturbations of the nonlocal RT gravity with nonzero background $S_i$. Conventions: the Greek indices run from 0 to 3, and the Latin indices run from 1 to 3. All numerical calculations in this paper are performed in the SI units.

\section{Cosmic background evolution}\label{sec:02}
The localized form of Eq. (\ref{eq:1.1}) can be written as \cite{Dirian2014,Kehagias2014,Belgacem2019a}
\begin{subequations}\label{eq:2.1}
\begin{align}
  G_{\mu\nu}+\frac{m^2}{6}(2Ug_{\mu\nu}+\nabla_\mu S_\nu+\nabla_\nu S_\mu)&=\kappa T_{\mu\nu},\label{eq:2.1a}\\
  \Box U&=-R,\label{eq:2.1b}\\
  \Box S_\mu+\nabla^\nu\nabla_\mu S_\nu&=-2\p_\mu U,\label{eq:2.1c}
\end{align}
\end{subequations}
where $U$ and $S_\mu$ are the auxiliary fields. One can directly verify that energy and momentum conservation can be derived from the above equations. In order to be consistent with current observations \cite{Aghanim2018} and the inflation theory \cite{Guth1981}, we assume the universe is described by the flat Friedmann-Lema\^{i}tre-Robertson-Walker (FLRW) metric
\begin{equation}
  \dx s^2=-c^2\dx t^2+a^2\dx\mathbf{r}^2,
\end{equation}
where $a=a(t)$. For the perfect fluid, we know the energy-momentum tensor $T^\mu_{\ \nu}={\rm diag}(-\rho c^2,p,p,p)$. For the auxiliary fields, we assume $U=U_0(t)$ and
\begin{equation}
  S_\mu=(c^2\mathcal{S}_0,a\mathcal{S}_1,a\mathcal{S}_1,a\mathcal{S}_1),
\end{equation}
where $\mathcal{S}_0=\mathcal{S}_0(t)$ and $\mathcal{S}_1=\mathcal{S}_1(t)$. Here we set $S_i=a\mathcal{S}_1$ because the universe is isotropic. This is the core difference between our work and previous works (e.g., \cite{Maggiore2014a,Dirian2014,Dirian2015,Dirian2016,Foffa2014,Nesseris2014}) that assume $S_i=0$.

Substituting the above assumptions into Eq. (\ref{eq:2.1a}), the $0i$-component gives
\begin{equation}
  \dot{\mathcal{S}}_1-H\mathcal{S}_1=0,
\end{equation}
where $\dot{}\equiv\dx/\dx t$ and the Hubble parameter $H\equiv\dot{a}/a$. Integrating the above equation gives
\begin{equation}
  \mathcal{S}_1(t)=l_1\frac{a(t)}{a_1},
\end{equation}
where $a_1$ can be regarded as the value of $a(t)$ at one specific time point, and $l_1$ is the integral constant with dimension of length [see Eq. (\ref{eq:3.3}) for the dimension]. In principle, $l_1/a_1$ is just one parameter. However, in order to facilitate the dimensional analysis of the following calculations, we reserve these two parameters. This solution means $S_i\propto a^2$, i.e., the value of $S_i$ increases as the universe expands. In other words, $S_i=0$ that used in \cite{Maggiore2014a,Dirian2014,Dirian2015,Dirian2016,Foffa2014,Nesseris2014} is unstable. Taking into account the above solution of $\mathcal{S}_1$, Eq. (\ref{eq:2.1}) gives
\begin{subequations}\label{eq:2.6}
\begin{gather}
  \frac{m^2}{3}U_0-\frac{m^2}{3}\dot{\mathcal{S}}_0-\frac{3H^2}{c^2}=-\kappa\rho c^2,\\
  \frac{m^2}{3}U_0-\frac{m^2}{3}H\mathcal{S}_0-\frac{2\ddot{a}}{c^2a}-\frac{H^2}{c^2}=\kappa p,\\
  \ddot{U}_0+3H\dot{U}_0=6\frac{\ddot{a}}{a}+6H^2,\\
  \ddot{\mathcal{S}}_0+3H\dot{\mathcal{S}}_0-3H^2\mathcal{S}_0=\dot{U}_0,
\end{gather}
\end{subequations}
which determine the evolution of the universe. Eq. (\ref{eq:2.6}) shows $U_0$ is dimensionless and the dimension of $\mathcal{S}_0$ is time. The surprising thing is that $l_1$ does not appear in Eq. (\ref{eq:2.6}), which means the cosmic background evolution in the nonlocal RT gravity is independent of $S_i$. In the following sections, we study the influence of the nonzero background $S_i$ on the scalar and tensor perturbations.

\section{Scalar perturbation}
In this section, we analyze the scalar perturbation of the nonlocal RT gravity with nonzero background $S_i$. Especially, we focus on the Newtonian approximation. The perturbed metric can be written as
\begin{equation}
  \dx s^2=-c^2(1+2\varepsilon\Phi/c^2)\dx t^2+a^2(1-2\varepsilon\Psi/c^2)\dx\mathbf{r}^2,
\end{equation}
where $\Phi=\Phi(\mathbf{r},t)$ and $\Psi=\Psi(\mathbf{r},t)$. Here and hereafter we use $\varepsilon$ to denote the first-order perturbation, and we set $\varepsilon=1$ after the Taylor expansion. Note that the FLRW background is necessary to clarify the possible time-varying $G$ \cite{Belgacem2019a,Tian2019a}. For the matter, the only nonzero component of $T_{\mu\nu}$ at the first-order is $T_{00}=\varepsilon\rho c^4$ \cite{Tian2019a}. For the auxiliary fields, we assume
\begin{subequations}\label{eq:3.2}
\begin{align}
  U&=U_0(t)+\varepsilon U_1(\mathbf{r},t),\\
  S_0&=c^2\mathcal{S}_0(t)+\varepsilon c^2\xi_0(\mathbf{r},t),\\
  S_i&=\frac{l_1a^2}{a_1}+\varepsilon a\xi_i(\mathbf{r},t),\label{eq:3.2c}
\end{align}
\end{subequations}
and then the dimensions of $\xi_0$ and $\xi_i$ are time and length, respectively.

Substituting the above assumptions into Eq. (\ref{eq:2.1a}), the $ij\,(i\neq j)$-component gives
\begin{align}\label{eq:3.3}
  &\frac{\p^2(\Psi-\Phi)}{\p x^i\p x^j}
  +\frac{m^2c^2a}{6}\frac{\p}{\p x^i}\left(\xi_j+\frac{2l_1a}{a_1}\frac{\Psi}{c^2}\right)\nonumber\\
  &\quad+\frac{m^2c^2a}{6}\frac{\p}{\p x^j}\left(\xi_i+\frac{2l_1a}{a_1}\frac{\Psi}{c^2}\right)=0.
\end{align}
Integrating the above equation gives
\begin{align}
  \xi_i&=-\frac{2l_1a}{a_1}\frac{\Psi}{c^2}+\frac{6}{m^2c^2a}\frac{\p\zeta}{\p x^i},\label{eq:3.4}\\
  \Psi&=\Phi-2\zeta,\label{eq:3.5}
\end{align}
where $\zeta=\zeta(\mathbf{r},t)$ is an arbitrary function with dimension of the square of speed. Taking into account the above solutions, the $ii$-component of Eq. (\ref{eq:2.1a}) gives
\begin{equation}\label{eq:3.6}
  \frac{2}{a^2}\nabla^2\zeta+\frac{\gamma m}{3a}\tilde{\nabla}(2\zeta-\Phi)+\mathcal{O}(m^2\Phi)=0,
\end{equation}
and the $00$-component gives
\begin{equation}\label{eq:3.7}
  \frac{2}{a^2}\nabla^2(\Phi-2\zeta)-\frac{\gamma m}{3a}\tilde{\nabla}\Phi+\mathcal{O}(m^2\Phi)=\kappa\rho c^4,
\end{equation}
where the dimensionless parameter $\gamma\equiv aml_1/a_1$ and the differential operator $\tilde{\nabla}=\p/\p x+\p/\p y+\p/\p z$. In the next section, based on the gravitational wave (GW) observations, we show that it is reasonable to assume $|\gamma|\lesssim\mathcal{O}(1)$ at today. For the Newtonian approximation, we have $m\ll\p/(a\p x)$ \cite{Tian2019a}. Thus the leading term of Eq. (\ref{eq:3.6}) gives $\nabla^2\zeta=0$ and the leading term of Eq. (\ref{eq:3.7}) gives the Poisson equation if $\kappa=8\pi G/c^4$. Without loss of generality, we can set $\zeta=0$, which gives $\Psi=\Phi$ as observations required \cite{Tian2019a}. Note that a constant but nonzero $\zeta$ can be absorbed by re-scaling the spatial coordinates. In summary, even considering the nonzero background $S_i$, the nonlocal RT gravity can still give the desired Poisson equation, $\Psi=\Phi$ and time-independent $G$ in the weak field limit (see \cite{Belgacem2019a,Tian2019a} for the observational constraints).

\cite{Kehagias2014} analyzed the spherically symmetric static solution of the nonlocal RT gravity with vanishing background $S_i$, and found the corrections to the Schwarzschild metric are of the form $1+\mathcal{O}(m^2r^2)$ in the region $r\ll m^{-1}$. However, as we will see, the appearance of the background $S_i$ would change this conclusion. In the region $r_{\rm S}\ll r\ll m^{-1}$, omitting the $\mathcal{O}(m^2\Phi)$ terms in Eqs. (\ref{eq:3.6}) and (\ref{eq:3.7}), we obtain the solutions
\begin{align}
  \frac{\Phi}{c^2}&=-\frac{r_{\rm S}}{2r}\left[1+\frac{\gamma m}{4}(x+y+z)\right],\\
  \frac{\zeta}{c^2}&=-\frac{r_{\rm S}}{24r}\,\gamma m(x+y+z),
\end{align}
and then Eq. (\ref{eq:3.5}) gives
\begin{equation}
  \frac{\Psi}{c^2}=-\frac{r_{\rm S}}{2r}\left[1+\frac{\gamma m}{12}(x+y+z)\right],
\end{equation}
where $r_{\rm S}=2GM/c^2$ is the Schwarzschild radius. Thus the leading term of the correction is $\mathcal{O}(mr)$ instead of $\mathcal{O}(m^2r^2)$. However, such correction is still unobservable. In addition, the above solutions show that the spacetime around the point mass is not spherically symmetric if $S_i\neq0$ in the background.

The above results are applicable to the solar system and binary star systems, but not to the cosmic large-scale structures, as we omit the $\mathcal{O}(m^2\Phi)$ term after Eq. (\ref{eq:3.6}). For the observational constraints involving the cosmological scalar perturbations, we should set $l_1/a_1$ as a parameter to fit the data. To do this, we need to modify the Einstein-Boltzmann solver as did in \cite{Dirian2014,Dirian2015,Dirian2016}. And we would like to leave the work to the future. However, one important thing is worth mentioning here. Eqs. (\ref{eq:3.2c}) and (\ref{eq:3.4}) show the perturbation of $S_i$ is proportional to $a^2\Psi$, which is similar to the behavior of the background $S_i$. But we do not think this growth will cause a fatal blow to the theory. The reason is $\delta S_i\propto a^2$ even if $l_1=0$ (see Pages 11, 29 and 30 in \cite{Dirian2014}), and this case can indeed fit observations well \cite{Dirian2014,Dirian2015,Dirian2016}. Nonzero $l_1$ can change the value of $\delta S_i$ only by the same order of magnitude if $l_1$ is not extremely large \footnote{Dimensional analysis of Eq. (\ref{eq:3.4}) indicates we should require $l_1a/a_1\lesssim\mathcal{O}(m^{-1})$, which is equivalent to $\gamma\lesssim\mathcal{O}(1)$ obtained in the next section.}. Therefore, it is reasonable to believe that observations allow the existence of nonzero $l_1$.

\section{Tensor perturbation}
In this section, we analyze the GW propagation in the nonlocal RT gravity with nonzero background $S_i$. Since observations prefer pure tensor modes than pure vector or scalar modes (see GW170814 \cite{Abbott2017_170814} and GW170817 \cite{Abbott2019_170817GR} for examples), here we only consider the tensor modes. Without loss of generality, we assume GW propagates in the $z$-direction. The perturbed metric can be written as
\begin{subequations}
\begin{equation}
  \dx s^2=-c^2\dx t^2+g_{ij}\dx x^i\dx x^j,
\end{equation}
where
\begin{equation}
  g_{ij} = a^2
  \left( \begin{array}{ccc}
  1+\varepsilon h_+ & \varepsilon h_\times & 0 \\
  \varepsilon h_\times & 1-\varepsilon h_+ & 0 \\
  0 & 0 & 1
  \end{array} \right),
\end{equation}
\end{subequations}
and $h_+=h_+(z,t)$, $h_\times=h_\times(z,t)$. For the energy-momentum tensor, all the components vanish at $\mathcal{O}(\varepsilon)$-order. For the auxiliary fields, we also assume Eq. (\ref{eq:3.2}) but replace $\mathbf{r}$ with $z$.

Substituting the above assumptions into Eq. (\ref{eq:2.1a}), we obtain \footnote{Firstly, we rise the $\mu$-index in Eq. (\ref{eq:2.1a}). Then the $12$-component gives the evolution equation of $h_\times$, and the difference between $11$ and $22$-components gives the evolution equation of $h_+$.}
\begin{equation}\label{eq:4.2}
  \frac{\p^2h}{\p t^2}+(3+\alpha)H\frac{\p h}{\p t}-\frac{c^2}{a^2}\frac{\p^2h}{\p z^2}+\frac{m^2c^2l_1}{3a_1}\frac{\p h}{\p z}=0,
\end{equation}
where the dimensionless parameter $\alpha=-m^2c^2\mathcal{S}_0/(3H)$. Here we omit the subscripts because $h_+$ and $h_\times$ satisfy the same evolution equation. Hereafter we define the dimensionless constant $\gamma\equiv a_3ml_1/a_1$, where $a_3$ is the value of the scale factor at today. Note that this definition is slightly different from the definition in the previous section. Using the Fourier transformation $h(z,t)=\int_{-\infty}^{+\infty}\hat{h}(k,t)e^{ikz}\dx k$, we obtain
\begin{equation}\label{eq:4.3}
  \ddot{\hat{h}}+(3+\alpha)H\dot{\hat{h}}+\left(\frac{c^2k^2}{a^2}+\frac{i\gamma kmc^2}{3a_3}\right)\hat{h}=0.
\end{equation}
If we ignore the $\dot{\hat{h}}$ term and assume $a=a_3$, then the solution of Eq. (\ref{eq:4.3}) is $\hat{h}=c_1e^{i\omega t}$, where $c_1$ is the integral constant and
\begin{equation}\label{eq:4.4}
  \omega=\pm\sqrt{\frac{c^2k^2}{a_3^2}+\frac{i\gamma kmc^2}{3a_3}}.
\end{equation}
Without loss of generality, in the following, we assume $k>0$ and adopt the minus sign in Eq. (\ref{eq:4.4}), which corresponds GW propagates along the positive direction of the $z$-axis. For the GWs detected by the ground-based detectors, we have $a_3/k\ll m^{-1}$, i.e., the wavelength is much shorter than the cosmic scale. This relation allows us to take Taylor expansion for Eq. (\ref{eq:4.4}), which gives
\begin{equation}\label{eq:4.5}
  \omega=-\frac{ck}{a_3}(1+\frac{i\gamma ma_3}{6k}+\frac{\gamma^2a_3^2m^2}{72k^2}+\cdots).
\end{equation}
$\gamma^2$-term affects the GW dispersion relation, and the GW velocity is
\begin{equation}\label{eq:4.6}
  v=(1+\frac{\gamma^2a_3^2m^2}{72k^2})c.
\end{equation}
For the LIGO GWs, the typical frequency is $100\,{\rm Hz}$ and the typical wavelength $\lambda\approx3\times10^6\,{\rm m}$, which gives the typical wavenumber $k/a_3=2\pi/\lambda\approx2\times10^{-6}\,{\rm m}^{-1}$. For the parameter $m$, \cite{Maggiore2014a} gives $m\approx0.67H_0/c\approx5\times10^{-27}\,{\rm m}^{-1}$. Thus the typical value of $a_3^2m^2/k^2$ is $6\times10^{-42}$. If $\gamma$ is not extremely large, Eq. (\ref{eq:4.6}) with such tiny value only leaves negligible effects in the GW dispersion relation and absolute velocity measurements. For example, GW170817 and GRB 170817A give $v/c=1\pm\mathcal{O}(10^{-15})$ \cite{Abbott2017_GW-GRB}, which requires $\gamma^2a_3^2m^2/(72k^2)<10^{-15}$, i.e., $\gamma<10^{14}$. In other words, the modification that appears in Eq. (\ref{eq:4.6}) is unobservable with current observations if $\gamma=\mathcal{O}(1)$. $\gamma$-term affects the GW amplitude. This effect is independent of $k$, which is similar to the role of $\alpha$ in Eq. (\ref{eq:4.3}). Based on Eq. (\ref{eq:4.5}), we know if the propagation time is comparable to $1/H_0$, then this effect is observable. One important thing worth mentioning is that Eq. (\ref{eq:4.2}) shows the $\p h/\p z$ term could appear in the GW propagation equation, which extends the general propagation equation that used in the previous works (see \cite{Saltas2014,Nishizawa2018} for examples).

In order to quantify the impact of the $\p h/\p z$ term on the GW amplitude, we can no longer assume $a=a_3$ in Eq. (\ref{eq:4.3}). Here we assume the GW signal was emitted at $t=t_2$ and $a(t_2)=a_2$, and was detected at $t=t_3$ and $a(t_3)=a_3$. The relation between the redshift and scale factor is $1+z_{\rm red}=a_3/a_2$. Taking the coordinate transformation $\eta(t)=\int_{t_2}^t\frac{a_2}{a(t')}\dx t'$, Eq. (\ref{eq:4.3}) can be written as
\begin{equation}\label{eq:4.7}
  \hat{h}''+(2+\alpha)\mathcal{H}\hat{h}'+\left(\frac{c^2k^2}{a_2^2}+\frac{i\gamma kmc^2a^2}{3a_2^2a_3}\right)\hat{h}=0,
\end{equation}
where $'\equiv\dx/\dx\eta$ and $\mathcal{H}\equiv a'/a$. Generally $\eta$ is called as the conformal time. This transformation is used to eliminate the time dependence of the $c^2k^2/a^2$ term in Eq. (\ref{eq:4.3}). In order to eliminate the $\mathcal{H}\hat{h}'$ term in Eq. (\ref{eq:4.7}), we use the function transformation $\hat{h}(\eta)=f(\eta)\tilde{h}(\eta)$, where
\begin{equation}
  f(\eta)=\exp\left(-\frac{1}{2}\int_0^\eta[2+\alpha(\eta')]\mathcal{H}(\eta')\dx\eta'\right),
\end{equation}
and then we obtain
\begin{equation}\label{eq:4.9}
  \tilde{h}''+\left[\frac{c^2k^2}{a_2^2}+\frac{i\gamma kmc^2a^2}{3a_2^2a_3}+\mathcal{O}(\mathcal{H}^2)\right]\tilde{h}=0.
\end{equation}
$\mathcal{O}(\mathcal{H}^2)$ term acts as the mass term in the general dispersion relation, and is negligible for current observations \cite{deRham2017}. Omitting the $\mathcal{O}(\mathcal{H}^2)$ term, the solution of Eq. (\ref{eq:4.9}) can be written as $\tilde{h}(\eta)=g(\eta)\cdot\exp(-ikc\eta/a_2)$, where
\begin{equation}
  g(\eta)=\exp\left(\frac{\gamma cm}{6a_2a_3}\int_0^\eta a^2(\eta')\dx\eta'\right).
\end{equation}
Then the GW amplitude is proportional to $f(\eta)\cdot g(\eta)$. Hereafter we denote $\eta_3\equiv\eta(t_3)$. For the standard siren, the GW luminosity disntace is inversely proportional to the amplitude \cite{Schutz1986}. In addition, the GW luminosity distance in general relativity satisfies $d_L^{\rm(GR)}\propto1/(fg)|_{\alpha,\gamma=0}$. Therefore, the ratio of the luminosity distance between the nonlocal RT gravity and general relativity is
\begin{align}
  &\frac{d_L^{\rm (RT)}}{d_L^{\rm(GR)}}=\frac{f(\eta_3)\cdot g(\eta_3)|_{\alpha,\gamma=0}}{f(\eta_3)\cdot g(\eta_3)}\nonumber\\
  &=\exp\left(\int_0^{\eta_3}\left[\frac{\alpha(\eta')\mathcal{H}(\eta')}{2}-\frac{\gamma cma^2(\eta')}{6a_2a_3}\right]\dx\eta'\right).
\end{align}
Transforming to the redshift, we obtain
\begin{align}
  \frac{d_L^{\rm (RT)}(z_{\rm red})}{d_L^{\rm(GR)}(z_{\rm red})}&=\exp\left(\int_0^{z_{\rm red}}\left[\frac{\alpha(\tilde{z})}{2(1+\tilde{z})}\right.\right.\nonumber\\
  &\qquad\qquad\left.\left.-\frac{\gamma cm}{6H(\tilde{z})\cdot(1+\tilde{z})^2}\right]\dx\tilde{z}\right).
\end{align}
If $\gamma=0$, the above equation is equivalent to the result obtain in \cite{Tsujikawa2019}. GW170817 \cite{Abbott2019_170817H0} rules out the possibility of $|\gamma|\gg1$. Future high redshift GW observations \cite{Belgacem2018a,Belgacem2018b,Belgacem2019b,Belgacem2019c,Belgacem2019d,Belgacem2019e} will provide tighter constraints on $\gamma$. Therefore, it is reasonable to assume $|\gamma|\lesssim\mathcal{O}(1)$ now.

The above discussion focuses on sub-horizon modes. Here we discuss the influence of the $\p h/\p z$ term on the evolution of super-horizon modes (primordial GWs), which is related to the early universe (inflation). The starting point is Eq. (\ref{eq:4.3}). CMB observations are the main method to detect primordial GWs \cite{Hu2014}. The typical wavenumber observed through CMB measurements is $k/a_3\approx H_0/c$ \footnote{Strictly speaking, $k/a_3\approx10^4H_0/c$, where the factor $10^4$ is related to the angular resolution of the CMB measurements. However, this factor does not affect our following conclusion. Especially, Eq. (\ref{eq:4.15}) is independent of $k$.}. The ratio of the two terms appear in Eq. (\ref{eq:4.3}) is
\begin{align}
  \frac{i\gamma kmc^2}{3a_3}/\frac{c^2k^2}{a^2}=\frac{i\gamma a_3m}{3k}\cdot\frac{a^2}{a_3^2}
  \lesssim\frac{a_e^2}{a_3^2}\approx10^{-64},
\end{align}
where $a_e$ is the scale factor at the end of inflation, and we estimate $a_e/a_3$ with $a_e/a_3\approx2.7\,{\rm K}/T_{\rm P}$, where $T_{\rm P}$ is the Planck temperature. Such tiny value means the influence of the $\gamma$-term on the GW dispersion relation is negligible in the super-horizon case. For a rough estimate of the influence on GW amplitude, we ignore the $\dot{\hat{h}}$ term in Eq. (\ref{eq:4.3}) and assume $a=a_e$, then the solution of Eq. (\ref{eq:4.3}) is $\hat{h}=c_1e^{i\omega t}$, where $c_1$ is the integral constant and
\begin{align}
  \omega&=-\sqrt{\frac{c^2k^2}{a_e^2}+\frac{i\gamma kmc^2}{3a_3}}\nonumber\\
  &=-\frac{ck}{a_e}-\frac{i\gamma cm}{6}\cdot\frac{a_e}{a_3}+\mathcal{O}(\frac{a_e^2}{a_3^2}).
\end{align}
Therefore, the main factor induced by the $\gamma$-term is
\begin{equation}\label{eq:4.15}
  \exp(-i\frac{i\gamma cm}{6}\frac{a_e}{a_3}t)
  \approx\exp(0.1\gamma H_0t\cdot\frac{a_e}{a_3})
  \approx1+\gamma\cdot\mathcal{O}(10^{-89}),
\end{equation}
where we assume $t\approx10^5t_{\rm P}$ and $t_{\rm P}$ is the Planck time. The above result shows the influence of the $\gamma$-term on the GW amplitude is also negligible in the super-horizon case. \cite{Lin2016,Tian2020} discussed the influence of nonzero $\alpha$ on the initial conditions of perturbations given by inflation. Figure 7 in \cite{Belgacem2019b} shows $|\alpha|\ll1$ in the early universe for the nonlocal RT gravity. These results indicate the $\alpha$-term appears in Eq. (\ref{eq:4.3}) is negligible in the early universe. Our results show the $\gamma$-term is also negligible. In summary, in the early universe, the GW evolution in nonlocal RT gravity is the same as in general relativity.

\section{Conclusions}
In this paper, after realizing that the background value of the auxiliary field $S_i$ increases with the expansion of the universe, we reanalyze the scalar and tensor perturbations of the nonlocal RT gravity with nonzero background $S_i$. For the scalar perturbation, we find the leading term of the corrections to $\Phi$ and $\Psi/\Phi$ is of the order of $\mathcal{O}(mr)$ instead of $\mathcal{O}(m^2r^2)$ obtained in \cite{Kehagias2014}. However, these corrections are still unobservable and thus the nonlocal RT gravity can still recover all successes (Poisson equation, $\Psi=\Phi$ and $G={\rm const.}$) of general relativity in the solar system as concluded in \cite{Dirian2014,Nesseris2014,Belgacem2019a,Tian2019a}. For the tensor perturbation, we find the $\p h/\p x^i$ term appears in the GW propagation equation, which extends the general propagation equation that used in \cite{Saltas2014,Nishizawa2018}. Our calculations show that the influence of the $\p h/\p x^i$ term on the GW dispersion relation is negligible in both the early and late-time universe, but the influence on the LIGO GW amplitude and also luminosity distance is observable if the dimensionless constant $\gamma=\mathcal{O}(1)$.

\section*{Acknowledgements}
This work was supported by the National Natural Science Foundation of China under Grant No. 11633001 and the Strategic Priority Research Program of the Chinese Academy of Sciences, Grant No. XDB23000000.

%

\end{document}